\documentclass[12pt,fleqn]{article}

\usepackage{graphicx}
\usepackage{pdftricks}

\begin{document}

\title{\Large Thermodynamic curvature from the critical point to the triple point}

\author{
   George Ruppeiner\footnote{ruppeiner@ncf.edu}\\
   Division of Natural Sciences\\
   New College of Florida\\
   5800 Bay Shore Road\\
   Sarasota, Florida 34243-2109 }

\maketitle

\begin{abstract}
I evaluate the thermodynamic curvature $R$ for fourteen pure fluids along their liquid-vapor coexistence curves, from the critical point to the triple point, using thermodynamic input from the NIST Chemistry WebBook.  In this broad overview, $R$ is evaluated in both the coexisting liquid and vapor phases.  $R$ is an invariant whose magnitude $|R|$ is a measure of the size of mesoscopic organized structures in a fluid, and whose sign specifies whether intermolecular interactions are effectively attractive ($R<0$) or repulsive ($R>0$).  I discuss five principles for $R$ in pure fluids: 1) near the critical point, the attractive part of the interactions forms loose structures of size $|R|$ proportional to the correlation volume $\xi^3$, and sign of $R$ negative, 2) in the vapor phase, there are instances of compact clusters of size $|R|$ formed by the attractive part of the interactions and prevented from collapse by the repulsive part of the interactions, and sign of $R$ positive, 3) in the asymptotic critical point regime, the $R$'s in the coexisting liquid and vapor phases are equal to each other, i.e., commensurate, 4) outside the asymptotic critical point regime incommensurate $R$'s may be associated with metastability, and 5) the compact liquid phase has $|R|$ on the order of the volume of a molecule, with sign of $R$ negative for a liquidlike state held together by attractive interactions and sign of $R$ positive for a solidlike state held up by repulsive interactions.  These considerations amplify and extend the application of thermodynamic curvature in pure fluids.

\end{abstract}

\noindent
{\bf Suggested PACS Numbers}: 05.70.-a, 05.40.-a, 64.60.Bd

\section{INTRODUCTION}

\par
Statistical mechanics and thermodynamics are based naturally in different domains.  Statistical mechanics starts at the microscopic level with known intermolecular interaction potentials.  Averaging with the Gibbs-Boltzmann distribution leads from this microscopic domain to the macroscopic domain where the laws of thermodynamics apply \cite{Pathria, Landau}.  But there is an intermediate mesoscopic domain where essential fluctuation phenomena connected with phase transitions take place.  The theme of this paper is getting information in this difficult domain using the thermodynamic curvature $R$.

\par
At least conceptually, statistical mechanics allows us to "build up" from the microscopic level to calculate properties in this mesoscopic regime.  But this operation is generally difficult in practice.  Less conceptually natural is thermodynamic fluctuation theory \cite{Pathria, Landau}, in which we ''build down'' to the mesoscopic domain from the thermodynamic one.  But information gets lost in the averaging yielding thermodynamics from statistical mechanics, and the idea of getting some of this information back seems counterintuitive at first.  Here, I do this backtracking using $R$, which connects to intermolecular interactions \cite{Ruppeiner1995, Ruppeiner2010}.

\par
In this paper, I calculate $R$ along the liquid-vapor coexistence curve from the critical point to the triple point for fourteen pure fluids, in both the liquid and vapor phases.   A schematic fluid phase diagram is shown in Figure 1, where $(p,T)$ denote pressure and temperature, respectively, with subscripts ''c'' and ''t'' for critical and triple point properties, respectively.  The thermodynamic input for this broad survey calculation comes from the NIST Chemistry WebBook \cite{NIST}, based on fits of experimental data.  Although the contents of this data base will not always be optimal in any particular region, or necessarily contain the latest experimental data, the NIST Chemistry WebBook represents the state of the art in representing data for many fluids over a large span of thermodynamic states.  The calculations in this paper mark the logical first step in understanding the broad behavior of $R$ in pure fluids.

\par
This paper is arranged as follows.  First, I summarize the method of calculation of $R$ in a number of thermodynamic coordinates.  Second, I give an overview of the physical interpretation of $R$ in pure fluids.  This overview includes both what was known previously, and what was learned here.  Third, I give results for $R$ calculated in fourteen pure fluids with the NIST Chemistry WebBook.  Fourth, I have an Appendix with proofs of critical point properties of $R$.

\begin{figure}[!t]
\centering
\includegraphics[width=3.0in]{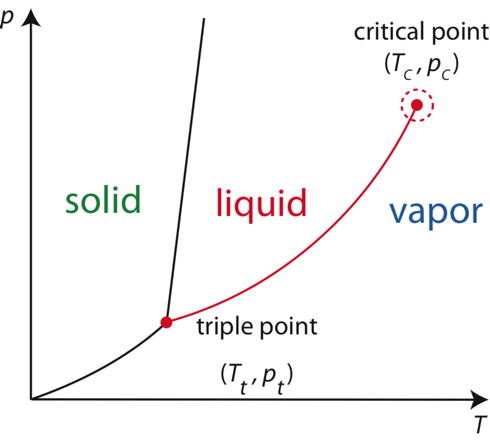}
\caption{A typical phase diagram for a material consisting of only one type of molecule.  Three phases, solid, liquid, and vapor are separated by first-order phase transition lines.  Of interest in this paper is the liquid-vapor coexistence curve connecting the triple point and the critical point.  The asymptotic critical region is indicated with an open circle.}
\end{figure}

\section{CALCULATION OF $R$}

\par
In this section, I summarize how $R$ is calculated in fluids in various coordinate systems.  For a given thermodynamic state, $R$ is an invariant, the same calculated in any thermodynamic coordinate system.  However, an appropriate coordinate system can much simplify a calculation.

\par
Consider a thermodynamic system consisting of one type of molecule, and with fundamental equation $U=U(S,N,V)$, where $U$ is the internal energy, $S$ is the entropy, $N$ is the number of molecules, and $V$ is the volume \cite{Callen}.  Define as well the temperature, chemical potential, and pressure: $\{T,\mu, p\}\equiv\{U_{,S},U_{,N},-U_{,V}\}$, where the comma notation indicates differentiation.  Defined in Table 1 are the Helmholtz free energy $A=A(T,N,V)$, the grand canonical potential $\Omega(T,\mu,V)$, and the Gibbs free energy $G = G(T,p,N)$.

\par
The thermodynamic entropy information metric $(\Delta\ell)^2$ is defined in terms of the fluctuation probability of an open subsystem with fixed volume $V$ of an infinite reservoir in a reference state 0 \cite{Pathria, Landau}:

\begin{equation}\mbox{probability}\propto\mbox{exp}\left[-\frac{V}{2}(\Delta\ell)^2 \right].\label{340}\end{equation}

\noindent $(\Delta\ell)^2$ is an invariant, positive definite quadratic form which in the pair of independent thermodynamic parameters $x^1$ and $x^2$ may be written as

\begin{equation}(\Delta \ell)^2\equiv g_{11}(\Delta x^1)^2 + 2 g_{12}\Delta x^1 \Delta x^2 +g_{22}(\Delta x^2)^2, \label{360}\end{equation}

\noindent where $\Delta x^\alpha\equiv (x^\alpha -x^\alpha_0)$ $(\alpha=1,2)$ denotes the difference between the thermodynamic parameters $x^{\alpha}$ of the subsystem and their values $x^{\alpha}_0$ corresponding to $(\Delta \ell)^2 =0$.  The thermodynamic metric elements $g_{\alpha\beta}$ are evaluated in the state $x^\alpha=x^\alpha_0$, and are tabulated below in Table 1 for several coordinate systems.

\par
The thermodynamic Riemannian curvature scalar (in the sign convention of Weinberg \cite{Weinberg}) may be written as\footnote{This equation is in footnote 53 of Ref. \cite{Ruppeiner1995}.  There is a small typographical error in Ref. \cite{Ruppeiner1995} as the term $\partial g_{12}/\partial x^2$ should be $\partial g_{12}/\partial x^1$.  The equation is given correctly in $(T,p)$ coordinates in Ref. \cite{Ruppeiner1990}.} \cite{Ruppeiner1995}

\begin{equation} \begin{array}{lr} {\displaystyle R= -\frac{1}{\sqrt{g}} \left[ \frac{\partial}{\partial x^1}\left(\frac{g_{12}}{g_{11}\sqrt{g}}\frac{\partial g_{11}}{\partial x^2}-\frac{1}{\sqrt{g}}\frac{\partial g_{22}}{\partial x^1}\right) \right. } \\  \hspace{3.6cm} + {\displaystyle \left. \frac{\partial}{\partial x^2}\left(\frac{2}{\sqrt{g}} \frac{\partial g_{12}}{\partial x^1} -\frac{1}{\sqrt{g}}\frac{\partial g_{11}}{\partial x^2}-\frac{g_{12}}{g_{11}\sqrt{g}}\frac{\partial g_{11}}{\partial x^1}\right)\right],}  \end{array} \label{380}\end{equation}

\noindent where

\begin{equation}g\equiv g_{11}g_{22}-g_{12}^2. \label{400}\end{equation}

\begin{table}
\begin{center}
\begin{tabular}{l|c|l}
\hline
\hline
$\{x^1,x^2\}$ & Potential & $\{g_{11},g_{12},g_{22}\}$ \\
\hline
$\{U,N\}$ & $S(U,N,V)$ & ${\displaystyle -\frac{1}{k_B V} \left\{\frac{\partial^2 S}{\partial U^2},\frac{\partial^2 S}{\partial U\,\partial N},\frac{\partial^2 S}{\partial N^2}\right\}}$ \\
$\{S,N\}$ & $U(S,N,V)$ & ${\displaystyle \frac{1}{k_B T V} \left\{\frac{\partial^2 U}{\partial S^2},\frac{\partial^2 U}{\partial S\,\partial N},\frac{\partial^2 U}{\partial N^2}\right\}}$ \\
$\{T,N\}$ & $A(T,N,V) = U - T S$ & ${\displaystyle \frac{1}{k_B T V} \left\{-\frac{\partial^2 A}{\partial T^2}, 0, \frac{\partial^2 A}{\partial N^2}\right\}}$ \\
$\{T,\mu\}$ & $\Omega(T,\mu,V) = U - T S - \mu N$ & ${\displaystyle -\frac{1}{k_B T V} \left\{\frac{\partial^2 \Omega}{\partial T^2},\frac{\partial^2 \Omega}{\partial T\,\partial\mu},\frac{\partial^2 \Omega}{\partial \mu^2}\right\}}$ \\
$\{T,p\}$ & $G(T,p,N) = U - T S + p V$ & ${\displaystyle -\frac{1}{k_B T V} \left\{\frac{\partial^2 G}{\partial T^2},\frac{\partial^2 G}{\partial T\,\partial p},\frac{\partial^2 G}{\partial p^2}\right\}}$ \\
\hline
\end{tabular}
\end{center}
\caption {Thermodynamic potentials and the thermodynamic metric elements in five coordinate systems.  Fluctuations for open subsystems take place at constant $V$, coinciding nicely with fixed $V$ in the derivatives in the first four coordinate systems.  In $\{T,p\}$ coordinates, however, $N$ is held fixed on differentiating $G(T,p,N)$.  Nevertheless, the form of the $\{T,p\}$ metric elements was picked to reflect fluctuations at constant $V$ \cite{Ruppeiner1990}.  Since $R$ is an invariant, the choice of coordinates for calculating it is purely one of convenience.}
\end{table}

\noindent $R$ is an intensive thermodynamic variable with units of volume per molecule.  Although the thermodynamic metric elements change their form on transforming coordinates, the value of $R$ for a given thermodynamic state does not change since it is an invariant, by the rules of Riemannian geometry.  For calculating $R$, the choice of coordinates is one purely of convenience.

\par
To conclude this section, I point out that Weinhold \cite{Weinhold} originated thermodynamic energy metrics in the form of inner products based on the Hessian of the internal energy. The positive-definite nature of these inner products represents the second law of thermodynamics.  But Weinhold's geometry lacks a true Riemannian metric structure since it has no underlying physical notion of distance, such as is offered by the fluctuation motivated entropy metric \cite{Ruppeiner1979}.  An entropy metric was also used by Andresen, Salamon, and Berry \cite{Andresen} as a measure of the dissipated availability in finite-time thermodynamics.  There have also been numerous calculations of $R$ for black hole thermodynamics; see \r{A}man et al. \cite{Aman} for review.

\section{PHYSICAL INTERPRETATION OF $R$}

\par
In this section, I present the basic thermodynamic curvature themes occurring in fluids.  The discussion here expands the physical interpretation of $R$ over what has been attempted previously.

\subsection{$|R|\propto\xi^3$ in the asymptotic critical region}

\par
For the single component ideal gas, $R=0$ whether the ideal gas is monatomic or molecular.  This finding originally motivated the hypothesis that $R$ measures intermolecular interactions \cite{Ruppeiner1979}.  The calculation of $R$ in model systems quickly revealed a proportionality between $|R|$ and the correlation volume: $|R|\propto\xi^d$, with $\xi$ the correlation length and $d$ the spatial dimensionality, particularly near the critical point where $\xi$ diverges \cite{Ruppeiner1995, Ruppeiner1979, Johnston2003, Brody2003}.  $|R|\propto\xi^d$ in the asymptotic critical region is consistent with all the fluid data examined in this paper.

\par
To evaluate the dimensionless proportionality constant between $|R|$ and $\xi^d$, model calculations in which both $R$ and $\xi$ are evaluated are required.  Several calculations of this type have been carried out in critical regions with large $\xi$: 1) four pure fluids near the critical point \cite{Ruppeiner1979}, 2) the one-dimensional ferromagnetic Ising model \cite{Ruppeiner1981}, and 3) the one-dimensional Takahashi gas \cite{Ruppeiner1990}.  In all these cases, the same proportionality constant was obtained:

\begin{equation} \xi^d=\frac{|R|}{2}.  \label{415}\end{equation}

\noindent $R<0$ in each case.  Based on these limited calculations, the proportionality constant $1/2$ would appear to be universal.  I am not aware of any proof of this universality, and this is a matter which would seem to deserve some future attention.

\par
Following this start based on model calculations, the connection $|R|\propto\xi^d$ was later confirmed on general grounds with a covariant thermodynamic fluctuation theory first developed by Ruppeiner \cite{Ruppeiner1983a, Ruppeiner1983b} and completed by Di\'osi and Luk\'acs \cite{Diosi} who explicitly added the conservation laws.  The idea is that at a large volume $V$ the Gaussian thermodynamic fluctuation theory Eq. (\ref{340}) works very well, but with decreasing $V$ the fluctuating subsystem eventually samples a correlated environment, which Eq. (\ref{340}) cannot model.  The covariant thermodynamic fluctuation theory predicts that Gaussian fluctuation theory ceases to work at a volume \cite{Ruppeiner1995,Ruppeiner2010}

\begin{equation} \tilde{V}\sim\frac{|R|}{6}.  \label{417}\end{equation}

\noindent In the asymptotic critical region, $\tilde{V}$ is physically interpreted as being roughly $\xi^d$.  Clearly, this interpretation is consistent with the result Eq. (\ref{415}) based on direct calculation.  For calculating the volumes of organized mesoscopic fluctuating structures below, I will use Eq. (\ref{417}) since its derivation is based on general arguments.

\par
A pictorial depiction of the meaning of the correlation length, due to Widom \cite{Widom1974}, is given in Figure 2.  Spontaneous density fluctuations cause the local density $\rho\left({\vec r}\right)$ at a point $\vec r$ in a single phase fluid to deviate from the overall density $\rho_0$ in some complex, time dependent manner. Mathematically, $\rho\left({\vec r}\right) = \rho_0$ corresponds to an intricate contour surface separating two sides with local mean densities $\rho\left({\vec r}\right) > \rho_0$ and $\rho\left({\vec r}\right) < \rho_0$.  A straight line through the fluid intersects this surface at points spaced an average distance $\xi$ apart.  $\xi$ is generally small in a disorganized system like an ideal gas, but diverges at the critical point for real fluids.  This figure also shows a ''droplet'' of linear dimension $\xi$ which offers schematic depictions of large spatially organized density fluctuations.  Such a schematic droplet near the critical point is shown in Figure 5a.

\begin{figure}[!t]
\centering
\includegraphics[width=3.0in]{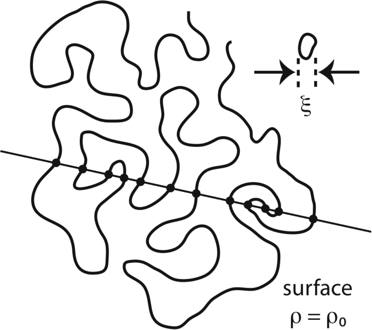}
\caption{A surface in three dimensions on which the local density $\rho\left({\vec r}\right)$ equals the overall density $\rho_0$.  Also shown is an arbitrary line intersecting the surface at the dotted points, and a ''droplet'' of linear dimension $\xi$ equal to the mean distance between those intersections.}
\end{figure}

\subsection{The sign of $R$}

\par
A tabulation of results in a number of systems \cite{Ruppeiner2010} suggests that the sign of $R$ reveals the basic character of the intermolecular interactions; $R$ is negative for thermodynamic states where attractive interactions dominate, and $R$ is positive for states where repulsive interactions dominate.  The clearest case is offered by the canonical examples of the Bose and Fermi ideal gasses, where quantum statistics causes atoms to either bunch closer together or further apart than in the corresponding classical ideal gas, mimicking the effect of attractive and repulsive interactions.  Janyszek and Mruga{\l}a \cite{Mrug90}, and Oshima, Obata, and Hara \cite{Oshima}, showed that (in Weinberg's sign convention \cite{Weinberg}) $R$ is always negative for the Bose gas and always positive for the Fermi gas.  Mirza and Mohammadzadeh \cite{Mirza2011} worked out the ideal $q$-deformed boson and fermion gasses, which also show the appropriate sign.  Brody and Ritz \cite{Brody2003} examined the sign of $R$ in finite Ising models as a function of the system size.

\par
Fluids are more complicated than ideal quantum gasses since the fluid intermolecular interaction potential typically has two parts, a repulsive part at short range and an attractive part at long range; see Figure 3.  Sorting out which part dominates in a given thermodynamic state can be difficult.  One statistical mechanical guide is offered by the pair correlation function

 \begin{equation} G(r)=\left[\frac{\rho(r)-\rho_0}{\rho_0}\right], \label{500}\end{equation}
 
\noindent where I have assumed that the position vector $\vec{r}$, with magnitude $r$, starts on a molecule within the bulk fluid.  Fisher and Widom \cite{Fisher1969} argued that attractive interactions dominate if the long-range decay of $G(r)$ is monotonic, and repulsive interactions dominate if this decay is oscillatory.  These authors \cite{Fisher1969} were the first to calculate the Fisher-Widom curve along which the long-range decay of $G(r)$ changes from monotonic to oscillatory.

\begin{figure}[!t]
\centering
\includegraphics[width=3.0in]{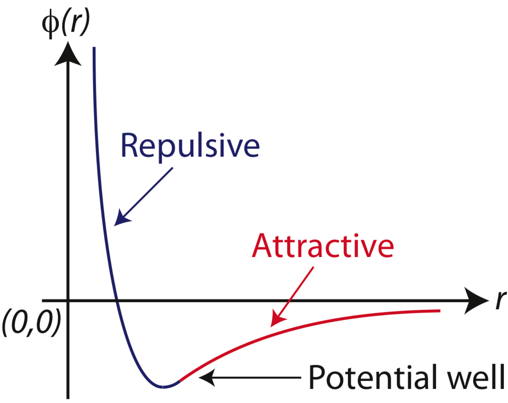}
\caption{Fluid intermolecular interaction potential $\phi(r)$ showing a repulsive force at short range, an attractive force at long range, and a potential well where molecules in the compact liquid or solid phases typically reside.}
\end{figure}

\par
It is natural to ask whether the Fisher-Widom curve might coincide with the curve $R=0$, since both these curves mark a transition from attractive to repulsive interactions.  Near the liquid-vapor critical point, the long-range decay of $G(r)$ is monotonic.  The basic lore is that attractive interactions dominate, and organize fluctuations over long distances.  One would thus expect negative $R$ in the asymptotic critical region, and this is found in all the fluids examined here.  As we move along the coexistence curve from the critical point towards the triple point, cases with $R$ switching sign from negative to positive are found with the NIST Chemistry WebBook \cite{NIST} in both the liquid and the vapor phases.  However, convincing results concerning the long-range decay of $G(r)$ are harder to come by.  The possible correspondence between the $R=0$ curve and the Fisher-Widom curve is under investigation.

\subsection{Commensurate $R$'s along the coexistence curve}

\par
Key in this research is the argument \cite{Ruppeiner2011} that the $R$'s in the coexisting liquid and vapor phases are equal to each other in the asymptotic critical region.  The structural underpinnings for this idea originated with Sahay, Sarkar, and Sengupta \cite{Sahay2010} who calculated $R$ for the van der Waals model, and observed that for isotherms with temperatures less than (but not too far from) the critical temperature, there must be an $R$-crossing point where the curvatures in the liquid and vapor phases equal each other.  This observation, augmented with a physical argument by Widom \cite{Widom1974}, led to the proposal by Ruppeiner et al. \cite{Ruppeiner2011} that this $R$-crossing point coincides with the coexistence curve.  In the Appendix, I offer a proof of this commensurate $R$ theorem in the context of the currently accepted asymptotic scaling description of fluid criticality.

\par
The physical argument \cite{Ruppeiner2011} for commensurate $R$'s envisions approaching some point on the coexistence curve from either the liquid phase or the vapor phase; see Figure 4, where $v$ is the molar volume.  Widom \cite{Widom1974} proposed that the correlation length in either of these bulk phases equals the thickness of the interface between the incipient coexisting phases.  Since this interface thickness is the same at some point on the coexistence curve, no matter from which direction we approach it, the correlation lengths, and hence the $R$'s, should be equal in the coexisting liquid and vapor phases.  This is the case even though the critical point joining the liquid and vapor phases is a singular point, and even though the coexisting liquid and vapor densities can be quite different from each other.  I add that Evans et al. \cite{Evans1993} showed with density functional theory using a short range intermolecular fluid potential that the character of the density decay, including the correlation length, at the interface matches that in the bulk, lending even further support to the Widom argument.

\par
This idea offers a practical new way of dealing with first-order phase transitions.  As is well known, $\xi$ can be difficult to calculate or to measure but $R$ follows readily from thermodynamic properties.  In the van der Waals model, the commensurate $R$ theorem was used in place of the physically problematic Maxwell equal area construction to derive the liquid-vapor coexistence curve \cite{Ruppeiner2011}.  May and Mausbach \cite{May2012} followed up by calculating the liquid-vapor coexistence curve for Lennard-Jones computer simulation data.  In both cases, strong results were obtained in this otherwise difficult problem.

\par
Widom \cite{Widom1974} also argued that if we have a majority phase of vapor on the verge of a first-order phase transition, then the fluctuating droplets within it have the density of the liquid phase which is to be made.  Likewise for the majority liquid phase preparing to make the vapor phase.  Figure 5c illustrates both this idea and the commensurate $R$ theorem.

\begin{figure}[!t]
\centering
\includegraphics[width=3.5in]{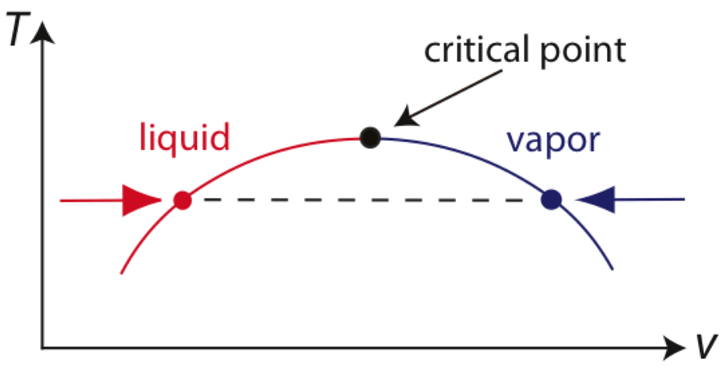}
\caption{The liquid-vapor coexistence curve and the critical point.  According to the commensurate $R$ theorem, at a given temperature $T$, the $R$'s in the two coexisting phases are equal, even though their molar volumes $v$ might differ considerably.}
\end{figure}

\begin{figure}[!t]
\centering
\includegraphics[width=3.0in]{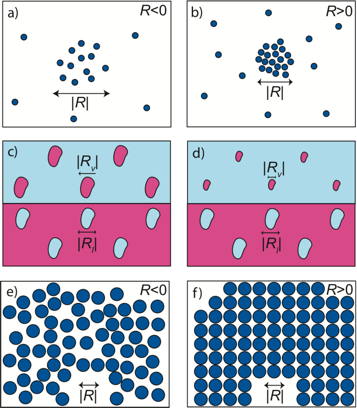}
\caption{Schematic figures illustrating several mesoscopic fluid concepts: a) a loose organization of molecules with volume $|R|$ pulled together by the attractive part of the intermolecular interactions ($R<0$), b) a compact cluster of molecules of volume $|R|$, pulled together by the attractive part of the intermolecular interactions, but prevented from collapse by the repulsive part of the intermolecular interactions ($R>0$), c) a fluid in two phases near the critical point, with the bottom half the liquid phase containing vapor droplets with volume $|R_l|$, under a coexisting commensurate vapor phase containing liquid droplets with the same volume $R_v=R_l$ as those in the liquid, d) liquid and vapor phases with incommensurate droplet sizes, e) a disorganized compact liquid phase held together by attractive intermolecular interactions ($R<0$), and f) an organized compact solid phase held up by the repulsive part of the intermolecular interactions ($R>0$).}
\end{figure}

\subsection{Low $|R|$ limit}

\par
Outside the asymptotic critical region, a limiting theme enters the picture \cite{Ruppeiner2011}.  With decreasing $|R|$, the correlation volume could become less than the molecular volume $v$, particularly in the vapor phase where $v$ typically becomes very large as the triple point is approached.  In this {\it low} $|R|$ {\it limit}, with $|R|/v\ll1$, we run short of statistics to do thermodynamics with at volume scales of $|R|$, and attempts at physical interpretation of $R$ might appear unreasonable.  Although such physical interpretations are attempted here anyway, the possibility that our use of $R$ might be stretched beyond its limits must always be considered.

\subsection{Incommensurate $R$'s along the coexistence curve}

\par
The commensurate $R$ picture associates liquid-vapor phase coexistence with mesoscopic structures of naturally the same spatial size $\xi$ in the two phases.  I conjecture that if these structures were of dissimilar sizes, as shown in Fig. 5d, and as is typically the case outside the asymptotic critical region, then it might be difficult for one phase to form the other, corresponding to metastability.

\par
Particularly difficult to treat theoretically is the onset of boiling as a liquid is warmed.  The classical homogeneous nucleation theory \cite{Brennen1995} is shown in Figure 6.  A bulk liquid at temperature $T$ contains a vapor bubble of radius $r$ and pressure $p_v$ attempting to expand against the sum of the liquid pressure $p_l$ and the pressure caused by the bubble's surface tension $\tilde{S}$ $(\tilde{S}>0)$.  It is generally assumed in this theory that all thermodynamic properties relate in the same way as in the thermodynamic limit, questionable at length scales approaching the order of intermolecular distances.

\par
For the bubble to grow, and form a macroscopic vapor phase, we require \cite{Brennen1995}

\begin{equation} p_v>p_l+\frac{2\tilde{S}}{r}, \label{520}\end{equation}

\noindent where $p_l$ is set by some external constraint, such as atmospheric pressure.  As the bulk liquid is warmed to the coexistence curve, $p_v(T)\to p_l$.  This limit cannot satisfy Eq. (\ref{520}), and, for the liquid to boil, we must superheat by warming to a $T$ higher than the phase transition point, with a larger $p_v$.

\begin{figure}[!t]
\centering
\includegraphics[width=3.0in]{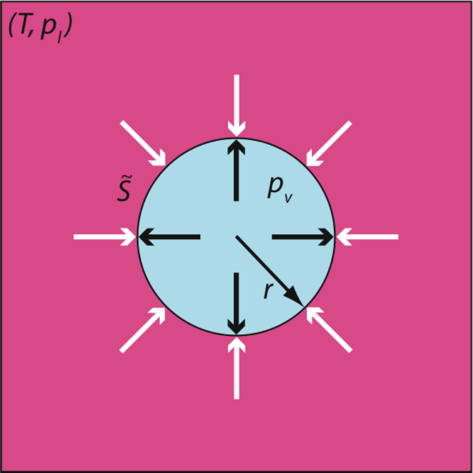}
\caption{Homogeneous nucleation theory picture of boiling.  A bubble of vapor with radius $r$ and pressure $p_v$ attempts to expand inside a liquid at temperature $T$.  To expand, the bubble must overcome the sum of the bulk liquid pressure $p_l$ and the pressure $2\tilde{S}/r$ contributed by the surface tension $\tilde{S}$.}
\end{figure}

\par
The homogeneous nucleation theory clearly cannot work for tiny bubbles, since $2\tilde{S}/r$ increases without limit as $r\rightarrow 0$.  Therefore, we usually assume that the bubbles are initially larger than some critical radius $r=r_c$ by forming on nucleation sites in the bulk liquid or on the container walls.  $\tilde{S}$ generally decreases to zero at the critical point \cite{Widom1974}, and so we would not expect much superheating in the asymptotic critical region.

\par
Temperatures below which superheating might become significant have been associated with corresponding isotherms having negative pressures in the metastable liquid regime.  In such a case, the liquid is able to withstand a pulling stress in the related cavitation problem \cite{Brennen1995}.  In simple fluid models, such negative pressures typically take place at $T<0.9 T_c$ \cite{Brennen1995} (for van der Waals, $T<27\,T_c/32$ \cite{Brody2009}).

\par
The homogeneous nucleation theory of boiling is conceptually challenged since, as becomes clear in section 4, for $T\sim 0.9\,T_c$, typically $|R|\sim 1$\,nm$^3$ in the liquid phase.  Such a volume contains roughly one vapor molecule, and it is hard to see how the macroscopic rules of thermodynamics assumed in this bubble model could possibly hold.  Zeng and Oxtoby \cite{Zeng} gave a critical analysis and concluded that ''although the standard classical theory has been quite successful in predicting droplet formation for many materials, it should fail completely in the process of bubble formation.''

\par
The suggestion in this paper that incommensurate $R$'s form a barrier to the onset of the liquid-vapor phase transition puts the focus on fluctuating mesoscopic structures, and without assuming anything about their particular character.

\subsection{$R$ in the compact liquid phase}

\par
The NIST Chemistry WebBook data presented in section 4 indicates that the compact liquid phase near the triple point has small $|R|$, on the order of the volume of a molecule.  The sign of $R$ is mostly negative; the few instances with positive $R$ coincide with having a curve with $R=0$ intersect the liquid phase.

\par
The only continuum fluid model where results for $R$ in the compact liquid regime have been reported is the one-dimensional Takahashi gas \cite{Ruppeiner1990}.  These results support small $|R|$ of the character found here.  I add to this calculation with a simple hard-sphere liquid model having Helmholtz free energy

\begin{equation} A(T, N, V) =N k_B T\, \mbox{ln}\left(\frac{N}{\tilde{\rho}\,V}\right) + N e(T) - N k_B T\,\mbox{ln}\left(1-\frac{b N}{V}\right),\label{540}\end{equation}

\noindent where $\tilde{\rho}$ is a constant with units of density, $e(T)$ is a function of $T$ with negative second derivative, and the constant $b$ gives the hard-packing limit.  The particle density is $\rho\equiv N/V$, the pressure $p=\rho k_B T/(1-b\rho)$, and the heat capacity at constant volume $C_v = -N Te''(T)$.  By Table 1 and Eq. (\ref{380}),

\begin{equation} R = -b(1-b\rho).\label{560} \end{equation}

\noindent Clearly, $-b\le R < 0$ in the regime of physical densities $0\le\rho < 1/b$.

\par
In the covariant version of the theory, $|R|$ sets the lower limit of applicability of thermodynamic fluctuation theory \cite{Ruppeiner1983a}.  Since for the hard-sphere gas, thermodynamic fluctuation theory is expected to work down to length scales on the order of the volume of a molecule $b$, the result $|R|\sim b$ is expected.

\par
This somewhat primitive model gives little insight into the sign of $R$, which in this model is uniformly negative.  To discuss the sign of $R$, I return to real fluids.  Near the critical point $R$ is always negative in the liquid state.  As I show with the real fluid data in section 4, as we cool the liquid state towards the triple point, $|R|$ typically decreases until it takes on a value of the order of the volume of a molecule.  Figure 5e pictures such a compact liquid state with $R<0$, loosely held together by attractive interactions.  We now have the question: as we cool further, could this state organize itself into a "solidlike" structure where hard-sphere repulsive interactions dominate and where $R$ changes sign to $R>0$, as shown schematically in Figure 5f?  As seen in section 4, such a sign change is indeed present in a few fluids.  I add that Widom \cite{Widom1967} contrasted the liquid behavior at the critical point and the triple point, and made the point that the former is dominated by long-range attractive interactions, and the later by short-range repulsive interactions.

\subsection{$R$ in the vapor phase}

\par
For the vapor phase $R$ is negative in the asymptotic critical region.  As we cool along the coexistence curve from the critical point to the triple point, we expect $|R|$ to decrease at first, and eventually end up negative near the triple point.  Negative $R$ is to be expected in any vapor regime where the molar volume is large, the molecules far apart, and the long-range attractive part of the intermolecular potential dominant over the repulsive part.  Negative $R$'s near the triple point are indeed seen in all the vapors considered in section 4.

\par
The magnitude of $R$ in the vapor phase at the triple point is, however, harder to interpret physically than its sign.  The fluids in section 4 show that either the vapor $R$ settles down to about $-1\,$nm$^3$ or decreases to large negative values.  The later case generally occurs in vapors with very large triple point molar volumes.  In either of these scenarios, the triple point $|R|/v\ll1$, and we are clearly below the length scale set by the low $|R|$ limit.  It would thus be easy to dismiss these vapor phase triple point results as being devoid of physical significance.  However, these results do not stand alone since qualitatively similar results were obtained for the one-dimensional Takahashi gas at large molecular volumes \cite{Ruppeiner1990}.  But I attempt no physical interpretation here.

\par
As we cool from the critical point, an interesting new feature presents itself in the vapor phases of a few fluids, such as water.  In water near the critical point, but outside the asymptotic critical region, $R$ crosses zero twice, with a peak at a positive value $R\sim +1\,$nm$^3$.  I associate the resulting regime of $R>0$ with the cluster forming scenario pictured in Figure 5b.

\section{NIST CHEMISTRY WEB BOOK RESULTS}

\par
In this section I report results of calculations of $R$ using fluid data from the NIST Chemistry WebBook \cite{NIST}.  The fluids I examined are made of molecules of four types: 1) monatomic molecules: Helium \cite{Ortiz2012}, Neon \cite{Katti1986}, Argon \cite{Tegeler1999}, Krypton \cite{Lemmon2006}, Xenon \cite{Lemmon2006}, and Methane \cite{Setzmann1991}, with Methane placed in this category because it is quasi-spherical \cite{Hansen2006}, 2) linear diatomic molecules: Normal Hydrogen \cite{Leachman}, Nitrogen \cite{Span2000}, and Oxygen \cite{Schmidt1985}, 3) other linear molecules: Carbon Dioxide \cite{Span1996} and Carbon Monoxide \cite{Lemmon2006}, and 4) more complicated molecules: Water \cite{Wagner2002}, Methanol \cite{deReuck1993}, and Hydrogen Sulfide \cite{Lemmon2006}.

\par
My calculations represent a broad based approach to significant themes for thermodynamic curvature in fluids.  Such a broad approach certainly fits the spirit of the NIST Chemistry WebBook, which is built on correlations, averaging, and extrapolations using numerous data sets in many fluids.  The NIST Chemistry WebBook is, however, not optimal in all circumstances.  For example, near the critical point it is not based on the scaled critical equations of state, and I will thus attempt no detailed critical point analysis, such as the evaluation of the critical point exponent of $R$.  But, fortunately, such analysis is not necessary in a first approach, given the critical point theorems presented in the Appendix.

\par
The basis of the calculation of $R$ is the Helmholtz free energy per volume

\begin{equation} f(T,\rho)\equiv \frac{A(T,N,V)}{V}, \label{570}\end{equation}

\noindent whose numerical values may be looked up in the NIST Chemistry WebBook \cite{NIST}.\footnote{Since the data tabulations are in quantities per mole, a multiplication by $\rho$ is required to convert to quantities per volume.}  The NIST Chemistry WebBook is based on fitting experimental data to smooth multiparameter formulas, yielding precise numbers very well suited to calculating numerical derivatives with finite difference formulas \cite{Abramowitz}.  In $(T,\rho)$ coordinates, Table 1 yields immediately the thermodynamic line element $\Delta\ell^2=g_{TT}\Delta T^2+g_{\rho\rho}\Delta\rho^2$, with

\begin{equation}\{g_{TT},g_{\rho\rho}\}=\left\{-\frac{1}{k_B T}\left(\frac{\partial^2 f}{\partial T^2}\right)_{\rho},     \frac{1}{k_B T}\left(\frac{\partial^2 f}{\partial\rho^2}\right)_T  \right\}.\label{600}\end{equation}

\noindent In these coordinates, Eq. (\ref{380}) becomes

\begin{equation}
R=\frac{1}{\sqrt{g}}\left[\frac{\partial}{\partial T} \left(\frac{1}{\sqrt{g}}\frac{\partial g_{\rho\rho}}{\partial T}\right)+\frac{\partial}{\partial \rho}\left(\frac{1}{\sqrt{g}}\frac{\partial g_{TT}}{\partial \rho}\right)\right],\label{620}
\end{equation}

\noindent with

\begin{equation} g\equiv g_{TT}\,g_{\rho\rho}. \label{625}\end{equation}

\par
Hydrogen offers a nice first case, with its thermodynamic properties recently reanalyzed \cite{Leachman}.  Figure 7a shows $R$ for Hydrogen along the coexistence curve in both the liquid and the vapor phases, from the critical point to the triple point.  A log-log plot of $R$ versus $(T_c-T)/T_c$ shows a power law divergence ($R\rightarrow-\infty$) at the critical point with slope near $2$ in both the liquid and vapor phases \cite{Ruppeiner2011}.  Such a divergence is seen at least approximately for all the fluids examined in this paper, and is expected from Eq. (\ref{2130}).  $R$ for the liquid and vapor phases agree with each other to better than $1\%$ in the temperature range $0.96<T/T_c<1$, consistent with the commensurate $R$ theorem.  By contrast, at $T/T_c=0.96$, the molar densities of the coexisting liquid and vapor phases differ from each other by a factor of $\sim 3$.

\newpage

\begin{figure}[t!]
\begin{minipage}[b]{0.5\linewidth}
\includegraphics[width=2.7in]{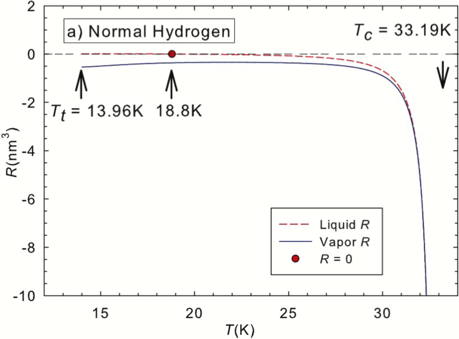}
\end{minipage}
\hspace{0.2cm}
\begin{minipage}[b]{0.5\linewidth}
\includegraphics[width=2.62in]{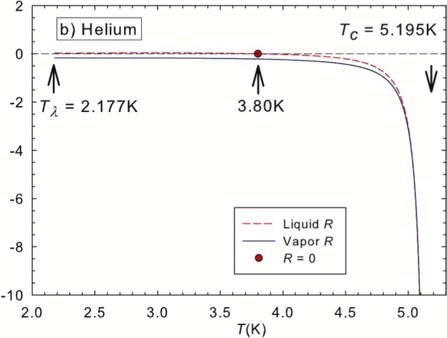}
\end{minipage}
\end{figure}

\begin{figure}[t!]
\begin{minipage}[b]{0.5\linewidth}
\includegraphics[width=2.7in]{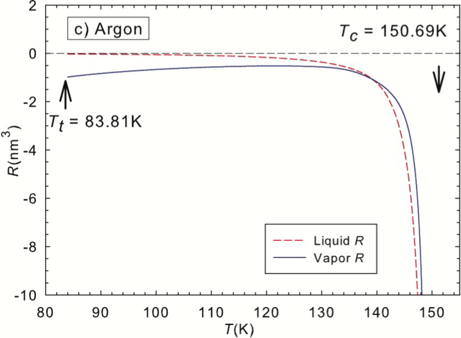}
\end{minipage}
\hspace{0.2cm}
\begin{minipage}[b]{0.5\linewidth}
\includegraphics[width=2.7in]{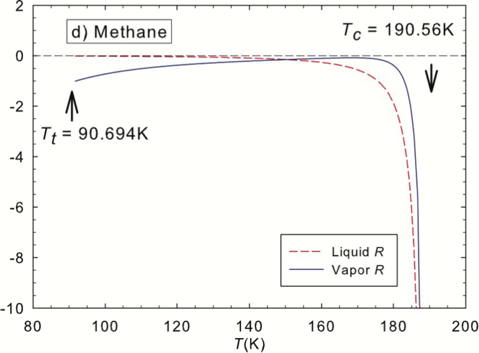}
\end{minipage}
\end{figure}

\begin{figure}[t!]
\begin{minipage}[b]{0.5\linewidth}
\includegraphics[width=2.7in]{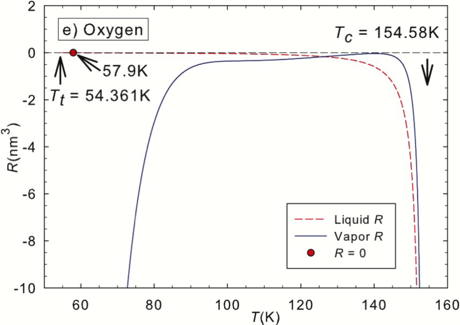}
\end{minipage}
\hspace{0.2cm}
\begin{minipage}[b]{0.5\linewidth}
\includegraphics[width=2.57in]{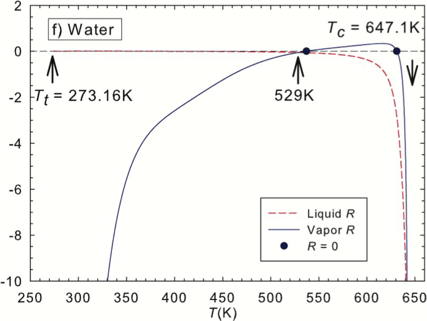}
\end{minipage}
\end{figure}
\noindent Figure 7:  $R$ for six representative fluids along their coexistence curves: a) Hydrogen, b) Helium, c) Argon, d) Methane, e) Oxygen, and f) Water.  The data go from the critical point, with temperature $T_c$, to the triple point, with temperature $T_t$ (except Helium which starts at the lambda point with temperature $T_\lambda $).  Each case shows $R$ diverging to $-\infty$ at the critical point, with the $R$'s more or less commensurate in the two phases in the asymptotic critical region.

\newpage
\begin{table}
\begin{center}
\footnotesize
\begin{tabular}{l|r|r|r|r|r|r|r}
\hline
\hline
Fluid & $T_t$ & $v_l$ & $v_v$ & $R_l$ & $R_v$& $R_l/v_l$ & $R_v/v_v$ \\
\hline
Neon & 24.56 & 0.01611 & 4.549 & -0.1654 & -0.379 & -6.182 & -0.050 \\
Argon & 83.81 & 0.02820 & 9.852 & -0.0259 & -0.983 & -0.552 & -0.060 \\
Krypton &115.78 & 0.03425 & 12.742 & -0.0286 & -1.697 & -0.502 & -0.080 \\
Xenon & 161.41 & 0.04426 & 15.962 & -0.0373 & -2.236 & -0.508 & -0.084 \\
Methane & 90.69 & 0.03553 & 63.925 & -0.0113 & -1.049 & -0.192 & -0.010 \\
\hline
Hydrogen & 13.96 & 0.02618 & 15.501 & +0.0083 & -0.5407 & +0.190 & -0.021 \\
Nitrogen & 63.15 & 0.03230 & 41.483 & -0.0162 & -0.7671 & -0.303 & -0.011 \\
Oxygen & 54.36 & 0.02450 & 3,081.0 & +0.0012 & -141.99 & +0.029 & -0.028 \\
\hline
CO$_2$ & 216.59 & 0.03735 & 3.1972 & -0.0694 & -0.6276 & -1.119 & -0.118 \\
CO & 68.16 & 0.03297 & 36.091 & -0.0177 & -0.9041 & -0.322 & -0.015 \\
\hline
Water & 273.16 & 0.01802 & 3,710.9 & -0.0018 & -114.01 & -0.059 & -0.019 \\
Methanol & 175.61 & 0.03542 & 7,834,000 & -0.0130 & -689,430 & -0.221 & -0.053 \\
H$_2$S & 187.70 & 0.03434 & 66.557 & -0.0155 & -1.2778 & -0.272 & -0.012 \\
\hline
\hline
\end{tabular}
\end{center}
\normalsize
\caption{Triple point table showing the triple point temperature $T_t$ in Kelvin, the molar volumes in the liquid and vapor phases, $v_l$ and $v_v$, in liters per mole, the liquid $R_l$ and the vapor $R_v$ in nm$^3$ per molecule, and $R_l/v_l$ and $R_v/v_v$ dimensionless.}
\end{table}

\par
For Hydrogen $|R|$ in the liquid phase decreases on cooling towards the triple point, with $R$ changing sign and remaining positive below $T=18.8$K.  At the triple point, $R=+8.3\times 10^{-3}\mbox{nm}^3$; see Table 2.  A corresponding sphere with volume $|R|/6$ would have radius 0.69\r{A}, of the same general order as the van der Waals radius of a Hydrogen atom 1.2\r{A}.  An organized structure with this size is in accord with the expectations from section 3.6 for the compact liquid phase.  As is clear from Table 2, such liquid phase triple point values of $|R|$ are typical for all the fluids.  Most of the liquid phase triple point $R$'s are negative, characteristic of compact disorganized structures, as in Fig. 5e, but there are three fluids with positive $R$, characteristic of the order in the solidlike states, as in Fig. 5f.

\par
In the vapor phase for Hydrogen, $R$ is uniformly negative.  This is reasonable, since in the asymptotic critical regime $R$ is negative, and as we cool towards the triple point the molecules become ever more widely spaced, and the attractive part of the interactions will thus increasingly gain in significance over the repulsive part.  Negative $R$ in the vapor phase is characteristic of all the fluids I looked at except Water and Methanol.  (Neon vapor has positive $R$ near the triple point, but this may be spurious.)

\par
$R$ for Helium, shown in Fig. 7b, looks qualitatively similar to that for Hydrogen, though the coexistence curve for Helium ends not at a triple point but at a lambda point at temperature $T_{\lambda}=2.177$K.  Again, in the asymptotic critical region, we see strong agreement with the commensurate $R$ rule.  In the vapor phase, $R$ is uniformly negative, but in the liquid phase $R$ changes sign and become positive below $T=3.80$K.

\par
$R$ for Argon, shown in Fig. 7c, has the commensurate $R$ theorem not as conspicuously satisfied as it was for Hydrogen and Helium.  Whether this reflects the real behavior, or shortcomings in the fitting formula in this regime is not clear.  For Argon, both the liquid and the vapor $R$'s are uniformly negative along the entire coexistence curve.  The basic pattern for Argon is repeated for Krypton, Xenon, Methane (see Fig. 7d), Nitrogen, Carbon Dioxide, Carbon Monoxide, and Hydrogen Sulfide.  Neon shows similar behavior as well, except for a regime of small positive $R$ in the liquid phase near the triple point.  Neon terminates in an anomalously large liquid $|R|/v$, as shown in Table 2.

\par
$R$ for Oxygen, shown in Fig. 7e, has the liquid phase terminate with a positive $R$ at the triple point.  $R$ changes sign at $T=57.9$K, with $|R|/v$ at the triple point anomalously small; see Table 2.  Whether or not this is a real effect, or reflects a problem with the data fit, is unclear.  The vapor phase shows a new feature, with $R$ decreasing abruptly on approaching the triple point.  As is indicated in Table 2, such large vapor phase $|R|$ values are typically associated with corresponding large increases in the vapor molar volume.  Water shows a similar feature, which may be related to metastability.

\par
$R$ for Water, shown in Fig. 7f, shows typical liquid phase behavior.  However, the vapor phase shows a new feature which may be physically significant, a {\it positive} peak for $R$ at $T=616$K, with $R=0.341$ nm$^3$.  (For this state $R/v=1.1$).  I associate this feature with cluster formation of the type shown in Fig. 5b, where water molecules are in a condensed solidlike state held up by the repulsive part of the intermolecular interactions.  It is straightforward to estimate the average number of water molecules in a cluster at the peak $R$.  By Eq. (\ref{417}), the cluster volume is roughly $|R|/6=0.0568$ nm$^3$.  Assume that each water molecule occupies a spherical volume with van der Waals radius: (4$\pi$/3)(0.17\mbox{nm})$^3=0.021\mbox{nm}^3$, leading to a number of molecules in the cluster of $0.0568/0.021\sim 3$ molecules.  Computer simulations by Johansson et al. \cite{Johansson2005} have found a significant number of water clusters in the coexisting vapor corresponding roughly to this peak $R$.  These clusters were found to be dominated by dimers, though clusters as large as 7 molecules were identified.

\par
Methanol shows a similar feature of positive $R$ in the vapor phase near the critical point, but the quality of the NIST fit may not be as good as for Water.  If this connection between positive $R$ and physical clusters is correct, then calculating $R$ offers a method of identifying fluids and thermodynamic regimes where we might find clusters.  This idea will be pursued elsewhere.

\par
For Water, there is a point at $T=529$K where the vapor curve for $R$ crosses the liquid curve, and a rapid divergence between the values of $R$ in the two phases occurs as we cool to the triple point.  If we interpret these large $R$ differences in terms of metastability, then we clearly see that, as we warm from the triple point, metastability should end at $T=529$K.  Brennen \cite{Brennen1995} gives the maximum temperature of superheating experimentally observed in Water as about 550K, with corresponding van der Waals value $T=27/32 T_c=546$K.

\section{CONCLUSION}

\par
In conclusion, this overview of the thermodynamic curvature $R$ on the liquid-vapor coexistence curve of fourteen pure fluids has confirmed some old expectations, but also brought forth some new features.  The broad trend is negative values of $R$, as the attractive part of the intermolecular interactions usually dominates for pure fluids, with the exception of compact situations where the molecules are in close contact.  In the asymptotic critical region, $|R|\propto\xi^3$ with the liquid and vapor values of $R$ in the coexisting phases equal by theorem.  On cooling towards the triple point, $|R|$ typically decreases until it is of the order of molecular sizes in the liquid phase, but less well defined in the vapor phase.

\par
Interesting new features are of three types.  The first is incommensurate $R$'s outside the asymptotic critical regime, with fluids of largely different $R$'s in the liquid and vapor phases.  I associate this with metastability, where the majority phase has a difficult time making the minority phase because the mesoscopic fluctuating structures in the phases are of different sizes.  Although I present no detailed experimental examination supporting this conjecture, it would appear to be at least logical.  Second, I associate isolated regimes of positive $R$ in the vapor phase near the critical point with the formation of solid clusters.  Third, fluids with $R$ changing sign through $R=0$ in the compact liquid phase may be associated with the Fisher-Widom curve marking the transition in the long-range decay of the correlation function from a monotonic fluidlike character to an oscillatory solidlike character.

\par
Generally, further study of $R$ might provide insight not only into the nature of thermodynamic curvature, but add another set of criteria to use in judging the quality of fits of fluid data to fitting models.

\section{ACKNOWLEDGEMENTS}

\par
I thank Peter Mausbach, Helge-Otmar May, Horst Meyer, Anurag Sahay, Tapobrata Sarkar, Guatam Sengupta, and Steven Shipman for useful communications.  Very special thanks to Eric Lemmon for programming $R$ into REFPROP 9.01 ($\beta$ test version).

\section{APPENDIX: COMMENSURATE $R$ THEOREM}

\par
Significant aspects of this paper hinge on the behavior of $R$ in the asymptotic critical region.  Below, I work in the context of the scaled form of the free energy, with the Rehr and Mermin \cite{Rehr} mixing of scaling variables to deal with asymmetry.  This constitutes the ''currently accepted asymptotic scaling description of fluid criticality'' \cite{Fisher2003}.  In this Appendix, I prove that $R$ has the same critical exponent as $\xi^3$ along the coexistence curve and its analytic extension into the supercritical regime.  I also prove the commensurate $R$ theorem that $R$ in the coexisting liquid and vapor phases are equal.

\par
My proofs assume that along the coexistence curve $\mu=\mu(T)$ is analytic approaching the critical point from below.  This assumption has been questioned in the context of the Yang-Yang anomaly where the second derivative $\mu''(T)$ diverges at the critical point.  There is some experimental evidence \cite{Fisher2000} for this point of view which requires ''complete scaling'' to address theoretically.  In this scenario, my proofs would require revisions.

\par
Write the pressure as \cite{Rehr}

\begin{equation}p(T,\mu)=p_0(T,\mu)+|\tau|^a Y_{\pm}(z)\label{2000},\end{equation}

\noindent where $p_0(T,\mu)$ is the regular analytic part, and the term containing the function $Y_{\pm}(z)$ is the singular part,

\begin{equation}z\equiv\frac{\zeta}{|\tau|^b},\label{2010}\end{equation}

\noindent $(a, b)$ are two critical exponents,

\begin{equation} \tau\equiv (T-T_c)/T_c + c_1(\mu-\mu_c)/k_B T_c, \label{2020}\end{equation}
 
\begin{equation} \zeta\equiv(\mu-\mu_c)/k_B T_c + c_2(T-T_c)/T_c ,\label{2030} \end{equation}

\noindent $(T_c,\mu_c)$ are the values of $(T,\mu)$ at the critical point, and $c_1$ and $c_2$ are two constants.  I assume $|c_1 c_2|<1$, the case if the fluid is not too antisymmetric.  The function $Y_{\pm}(z)$ has no explicit dependence on $c_1$ and $c_2$, and has two branches $(\pm)$ depending on the sign of $\tau$.  These branches join smoothly along the curve $\tau=0$, except at the critical point $\{\tau,\zeta\}=\{0,0\}$.  The critical exponents $a$ and $b$ are related to the standard critical exponents $\alpha$ and $\beta$ by $a=2-\alpha$ and $b=\beta\delta$ \cite{Rehr,Stanley}.

\par
It is assumed that $Y_{\pm}(z)$ is an even function of $z$.  This assumption, and the goal of modeling a first-order phase transition terminating in a second-order critical point, requires \cite{Rehr}:

\begin{equation} Y'_+(0)=0,\label{2040}\end{equation}
\begin{equation} Y^{(3)}_+(0)=0,\label{2050}\end{equation}
\begin{equation} Y_-(0_+)=Y_-(0_-)\ne 0,\label{2060}\end{equation}
\begin{equation} Y'_-(0_+)=-Y'_-(0_-)\ne 0,\label{2070}\end{equation}
\begin{equation} Y''_-(0_+)=Y''_-(0_-)\ne 0,\label{2080}\end{equation}

\noindent and

\begin{equation}Y^{(3)}_-(0_+)=-Y^{(3)}_-(0_-)\ne 0, \label{2090}\end{equation}

\noindent where the subscripts $+$ and $-$ on $0$ refer to the sign of $z$ as either $\zeta\rightarrow 0_+$ or $\zeta\rightarrow 0_-$.

\par
For $\tau<0$, $\zeta=0_{\pm}$ represents the two phases on the coexistence curve because, for given $(T, \mu$), we have $p(\tau,0_+)=p(\tau,0_-)$, by Eqs. (\ref{2000}) and (\ref{2060}).  This continuity of $(T,p,\mu)$, and the corresponding discontinuities of the density $\rho=(\partial p/\partial\mu)_T$ and entropy per volume $s=(\partial p/\partial T)_{\mu}$ resulting from Eqs. (\ref{2000}) and (\ref{2070}), are the conditions for a first-order phase transition.  Note that setting $\zeta=0_{\pm}$ in Eq. (\ref{2030}) yields $(\mu-\mu_c)/k_B T_c=-c_2\,t$ and $\tau=(1-c_1 c_2)t$ in Eq. (\ref{2020}), where the reduced temperature

\begin{equation}t\equiv\frac{T-T_c}{T_c}. \label{2100}\end{equation}

\noindent Thus, on the coexistence curve, $t$ is negative and $(\mu-\mu_c)/k_B T_c$ has the same sign as $c_2$.

\par
In $(T,\mu)$ coordinates the thermodynamic line element can be read off from Table 1:

\begin{equation} \Delta \ell^2 = \frac{1}{k_B T_c} \left[ \frac{\partial^2 p}{\partial T^2}\Delta T^2+2\frac{\partial^2 p}{\partial T\,\partial\mu} \Delta T \Delta\mu + \frac{\partial^2 p}{\partial \mu^2}\Delta\mu^2\right], \label{2110}\end{equation}

\noindent where I have set $\Omega=-p V$, and $T=T_c$ in the prefactor since we are in the asymptotic critical region.  It is straightforward to show that the term containing $Y_{\pm}(z)$ in Eq. (\ref{2000}) causes each second derivative in the thermodynamic line element Eq. (\ref{2110}) to diverge on approaching the critical point, assuming critical exponent values in the vicinity of the pure fluid ones ($a\sim 1.9, b\sim 1.6$ \cite{Stanley}).  Therefore, since the regular part of the pressure $p_0(T,\mu)$ produces only finite second derivatives in the metric elements, we can drop it in calculating $R$ in the asymptotic critical region.

\par
For $\tau>0$, and along the line $\zeta=0$ representing the analytic continuation of the coexistence curve, Eq. (\ref{380}) yields,

\begin{equation}
R_+(t,0)=
-\frac{
k_B T_c(b-1)(2b - a) (1-c_1 c_2)^{-a}
}
{a(a-1)Y_+(0)}\,t^{-a},
\label{2120}\end{equation}

\noindent demonstrating that  $R_+(t,0)$ diverges with exponent $a=2-\alpha$.  This exponent is the same as that of $\xi^3$ \cite{Stanley}.  To assure a positive heat capacity, we must have $Y_+(0)>0$, yielding $R_+(t,0)<0$ for fluids.  This is found in every case considered in this paper.

\par
Along the coexistence curve we have {\it exactly}

\begin{equation}
R_-(t,0_{\pm})=
\frac{
\begin{array}{lllr}
-k_B T_c(b-1)(1-c_1 c_2)^{-a}\\ \times[-(a-b)^2 \left(1+b-a\right)Y_-'(0_\pm)^2 Y_-''(0_\pm)\\ +\, 2 a (a-1) (2b-a) Y_-(0_\pm) Y_-''(0_\pm)^2\\ +\,a (a-1) (a-b) Y_-(0_\pm) Y_-'(0_\pm)Y_-^{(3)}(0_\pm)]\,|t|^{-a}
\end{array}
}
{2 \left[(a-b)^2 Y_-'(0_\pm)^2 - a (a-1) Y_-(0_\pm) Y_-''(0_\pm)\right]^2}.
\label{2130}\end{equation}

\noindent Every product of $Y_-(0_{\pm})$ and its derivatives has the same value in either phase by Eqs. (\ref{2060})-(\ref{2090}), establishing immediately the commensurate $R$ theorem: 

\begin{equation} R_-(t,0_-)=R_-(t,0_+).\label{2140}\end{equation}

\noindent Note also that $R_-$ diverges with exponent $a=2-\alpha$, the same as that of $\xi^3$.  The presence of the term $Y_-^{(3)}(0_\pm)$ in Eq. (\ref{2130}), with sign unset by thermodynamic stability, means that Eq. (\ref{2130}) does not clearly set the sign of $R_-(t,0_{\pm})$.  But it was found to be negative in the asymptotic critical region in all fluids examined in this paper.

\par
Other than $T$, $p$, $\mu$, and $R$, most thermodynamic functions will not be equal in the coexisting phases.  For example, the metric elements $g_{\alpha\beta}$, related to heat capacities and compressibilities, consist of a mixed sum of terms, some changing sign on switching phases and some not changing sign.

\end{document}